# Design of a 13-Channel Hybrid Array System for Foot/Ankle Magnetic Resonance Imaging at 7T/300MHz

Aditya Bhosale[1], Leslie Ying[1,2], Xiaoliang Zhang[1,2*]

[1]Department of Biomedical Engineering, State University of New York at Buffalo, Buffalo, NY 14260, USA

[2]Department of Electrical Engineering, State University of New York at Buffalo, Buffalo, NY 14260, USA

*Corresponding author, xzhang89@buffalo.edu

*Abstract*—Microstrip lines are being used in MR applications due to their unique properties, such as reduced radiation loss, high-frequency capability, and reduced perturbation of sample loading to the RF coil compared to conventional coils. Here, we present the design of the 13-channel hybrid array consisting of 12 Microstrips, 1 volume half birdcage coil placed on the foot/ankle phantom, and high permittivity materials to cover the maximum area of the subject at 7T/300MHz. We demonstrate using electromagnetic simulations, magnetic field distribution, SAR performance, and the coupling performance of the array elements. This work provides an ultrahigh field multichannel RF solution to lower extremity MR imaging with excellent imaging coverage and field uniformity.

## I. Introduction

**M**agnetic resonance imaging (MRI) [1, 2] is widely used to detect various abnormalities, injuries, and diseases in bones and soft tissues, such as cartilage degeneration, bone marrow edemas, osteoarthritis, osteoporosis, and ligament/tendon injuries [3-15]. The Magnetic resonance imaging modality has the upper hand in the early detection of abnormalities related to bones and soft tissues[16-26]. The Ultra-high field MRI imaging at static magnetic fields greater than or equal to 7T ( ≥300 MHz) offers increased Signal-to-noise ratio (SNR) and spatial and contrast resolution [27-43] but also has some drawbacks related to physiological side effects, increased specific absorption rate, challenges in RF hardware design, image degradation artifacts and heavy computations resulting from advanced data [10, 11, 39, 44-62]. Most MRI scanners used for MR foot/ankle imaging operate at 1.5 or 3 Tesla. The development of new RF systems and scanners operating at ultra-high field strength is increasing rapidly due to the promise of signal-to-noise ratio (SNR) gain [8, 57 , 63]. Given the critical role of RF transceiver arrays in SNR, irregular geometry of the anatomy, and complexity of RF magnetic field (B1) at ultrahigh fields, it is crucial to develop efficient RF coil array systems for foot/ankle imaging with sufficient B1 homogeneity and coverage in the specific area of interest while reducing the local specific absorption rate (SAR)[55, 64-68]. There are very few foot array coils available for UHF MR imaging, and the available array coils fail to produce uniform field distribution in the geometrically irregular human foot. To address the disadvantage mentioned above, we propose to design the 13-channel hybrid array system for foot imaging at 7T using mixed array elements of microstrip resonators and half-volume birdcage coil. Additionally, we used the high dielectric material's quality to manipulate the B1 field distributions to produce the uniform field distribution in the human foot-shaped phantom.

Microstrip transmission lines (MTL) are promising in high-frequency RF coil designs for MR applications at ultrahigh fields due to their unique advantages such as reduced radiation loss, distributed-element circuit, high-frequency capability, and reduced perturbation of sample loading to the RF coil in high-frequency range over conventional coils [58, 66, 69-83]. Microstrips are purely distributed and consist of a thin strip conductor, which can be silver or copper. Microstrip has a ground plane separated from the strip conductor by a low-loss dielectric material with a certain thickness. Microstrips provide a higher Q-factor and excellent decoupling performance over the conventional loop coils, require no shielding, have lower costs, and are easy to fabricate[44, 45, 77, 82, 84-88]. The frequency of the conventional MTL resonator can be calculated using the following equation [44, 45, 81, 85, 86, 89-91]:

$$f_r = \frac{nc}{2l\sqrt{\varepsilon_{eff}}}, (n = 1,2,3...).$$

The capacitively terminated MTL resonators are similar to the conventional MTL resonators. The termination capacitors can be connected to either one or both ends of the conductor. The termination capacitors increase the electrical length, making the microstrip line resonate at the desired frequency [44, 45, 82, 84-88, 92].

The following equation is used to calculate the effective dielectric constant of the MTL:

$$\varepsilon_{eff} = \left[1 + \frac{H_1 - H}{H}(\sqrt{\varepsilon_r} - 1)(\xi^+ - \xi^- \ln\frac{W}{H})\right]^2$$

Where

$$\xi^+ = \left(0.5173 - 0.1515 \ln\frac{H1 - H}{H}\right)^2$$

$$\xi^- = \left(0.3092 - 0.1047 \ln\frac{H1 - H}{H}\right)^2$$

The characteristic impedance of the MTL is calculated using the following equation:

$$z_0 = \frac{60}{\sqrt{\varepsilon eff}} \ln\left[\frac{H}{W}\right] \Xi + \sqrt{1 + (\frac{2H}{W})^2}$$

Where

$$\Xi = 6 + \frac{2(\pi - 3)}{exp\left(\frac{30.666H}{W}\right)^{0.7528}}$$

Where W is the width of the strip conductor, H is the height of the substrate

Finally, the resonant frequency of the MTL resonator terminated by one capacitor is calculated by using the following equation:

$$f_r = \frac{-1}{2\pi Z_0 C_t} \tan\left(\frac{2\pi l\sqrt{\varepsilon_{eff}}}{c} f_r\right)$$

And the resonant frequency of the MTL resonator terminated on both sides is calculated using the following equation:

$$f_r = \frac{(2\pi f_r Z_0)^2 C_t C_{t1} - 1}{2\pi Z_0 (C_t + C_{t1})} \tan\left(\frac{2\pi l\sqrt{\varepsilon_{eff}}}{c} f_r\right)$$

$Z_0$ is the characteristic impedance, $\varepsilon_{eff}$ is the effective permittivity, $l$ is the length of the strip conductor, and $C_t$ & $C_{t1}$ are the termination capacitors [45].

There are various types of RF coils that can be classified based on the working principle. One of the most commonly used RF coils is Birdcage volume coils. The birdcage coil provides excellent signal-to-noise ratio and B1 field uniformity and is considered safe for MRI applications. A birdcage coil consists of two circular rings called end rings and are connected by several legs or rungs of equal length. The end rings and legs are made of a conductive material, and the legs have capacitors connected to them. The birdcage coil can be tuned at its resonant frequency by selecting the appropriate capacitors [64, 93-96].

## II. METHODS

Twelve capacitively terminated microstrip transmission lines[44, 45, 84, 85] and one half-birdcage volume coil[64, 93-100] make up the proposed improved version. The dimensions and material properties of the human foot/ankle-shaped phantom were altered to match the human foot closely. We used eight 11.6 cm long microstrips arranged in a circular pattern over the ankle area of the human foot/ankle-shaped phantom. The other four microstrips were 20 cm long and placed over the metatarsal and phalanges regions of the phantom's human foot/ankle. We used a single half-birdcage coil to replace the surface coils and wrapped it around the heel of the human foot/ankle-shaped phantom. For all microstrips, the strip conductor width is 1 mm, the substrate width is 1 cm, and the substrate height is 1 cm. 7 rungs/legs and two end rings make up the half-birdcage coil. Each leg measures 9.2 centimeters in length and 1 centimeter in width, and the end rings are 4mm in diameter. Below the foot phantom, we added a 4 mm thick dielectric layer [101-105]. The simulation model's layout and design are shown in the following diagrams.

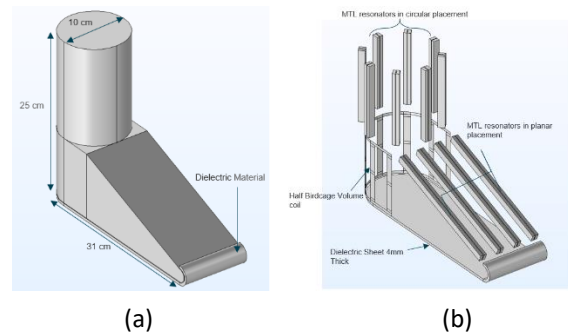

(a)      (b)

Fig. 1 (a) The Geometry of the modified Human foot/ankle-shaped phantom and dimensions. (b) The structure of the 13-channel hybrid array system.

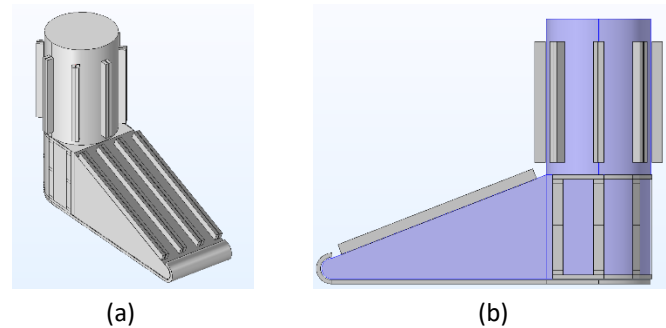

(a)      (b)

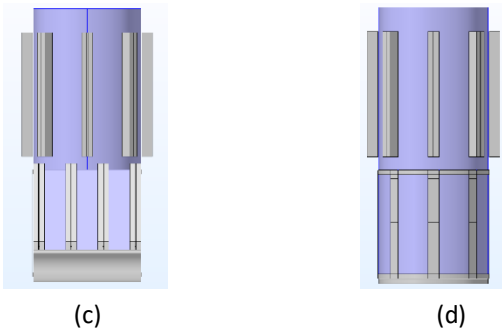

Fig. 2 The geometry of the 13-channel hybrid array system loaded with the human foot/ankle-shaped phantom (a) Three-dimensional view (b) Side view (c) Front view (d) Back view.

The capacitors were added using lumped elements, and each channel was excited using lumped ports. The circular arrangement of microstrip lines was excited with equal amplitude and a phase difference of $45^0$ for each microstrip line. The metatarsal and phalangeal microstrip lines were excited with the same amplitude with no phase difference. The half-birdcage coil's legs were excited with the same amplitude and a $45^0$ phase difference. Two 6.15pF termination capacitors were used to tune the 11.6 cm long Microstrip lines to 300MHz, and two 2.7pF termination capacitors were used to tune the 20cm long Microstrip lines at 300MHz. The half-birdcage coil uses 14 capacitors placed on both ends of each leg, and 8.85 pF capacitors were used to tune the half-birdcage coil at 300MHz. The capacitor values for the microstrip lines were tested using the Microstrip resonant frequency calculator. We evaluated the dielectric sheet's performance at the following relative permittivity values: 300, 500, 700, and 1000. The magnetic field distribution, decoupling performance, and SAR studies of the 13-channel hybrid array device were tested using electromagnetic simulations with and without the dielectric sheet.

*A. Substrate*

The material used as the substrate for the MTL resonator is commercially available as RO4003C Laminate. The dielectric material properties of the substrate were $\varepsilon_r$ = 3.38, σ = 0 s/m.

*B. Phantom*

In our simulation model, we used one human foot/ankle-shaped phantom. The dielectric material properties were set to $\varepsilon_r$ = 39, σ = 0.49 s/m to reflect the human foot/ankle properties. Figure 1 shows the model using the human foot/ankle-shaped phantom.

*C. High Dielectric Sheet*

We placed a high dielectric sheet with a thickness of 4mm below the foot phantom to evaluate the dielectric sheet's effect on the field distributions. The high dielectric sheet's relative permittivity was varied to test the array system's performance at different values. Figure 1 shows the dielectric sheet placed beneath the foot/ankle phantom.

*D. Simulations*

We performed electromagnetic simulations using COMSOL Multiphysics software. We computed the frequency domain method to compute our array system's frequency response at 7T/300 MHz. Coupling performance and Specific absorption rate were also calculated using simulation studies.

*F. Data Analysis*

We exported the B field maps, current density plots, and S-parameters directly using COMSOL Multiphysics software. The magnetic field distribution was displayed using the following expression:

$$20 \times \log_{10}(emw.normB)$$

The current density plots were constructed using the following expression:

$$emw.normJ$$

Finally, the S-parameters were evaluated using the following AWE expression:

$$abs(comp1.emw.S21)$$

III. RESULTS

*A. Decoupling Performance/ S-parameters Evaluation*

We tested the coupling between microstrip lines in a circular and planar configuration. We used two microstrip lines in this experiment, but only one was excited while the other was held as a passive element. In the circular arrangement, the gap between neighboring microstrip line elements was 2.7 cm, while in the planar configuration, it was 1.9 cm. The S11 and S21 parameters for circularly placed microstrip lines were -41.5 and -43 dB, respectively. In planar positioning, the S11 and S21 parameters for microstrip lines were -38 and -26 dB, respectively. Our findings demonstrate excellent decoupling between the Microstrip transmission line channels without using any decoupling techniques.

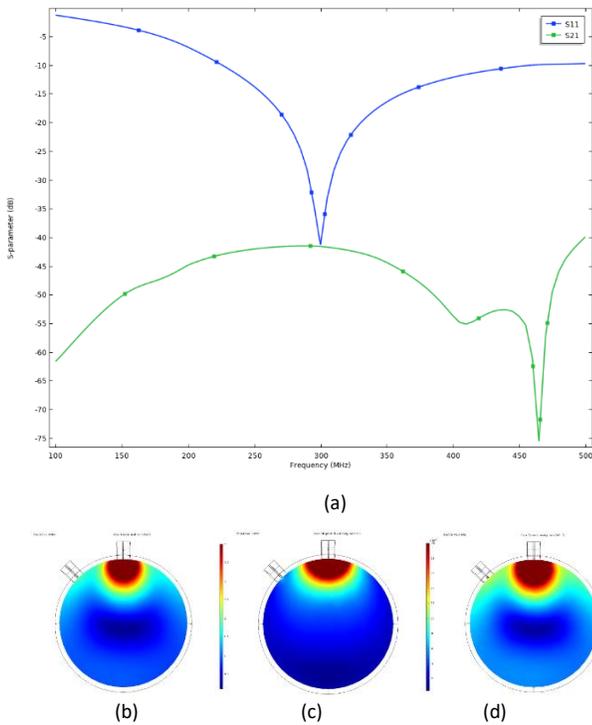

Fig. 3. The decoupling performance of the MTL resonators in a circular configuration. (a) S parameters evaluation (b) Electric fields, (c) magnetic fields, and (d) currents induced in the passive element from the active component.

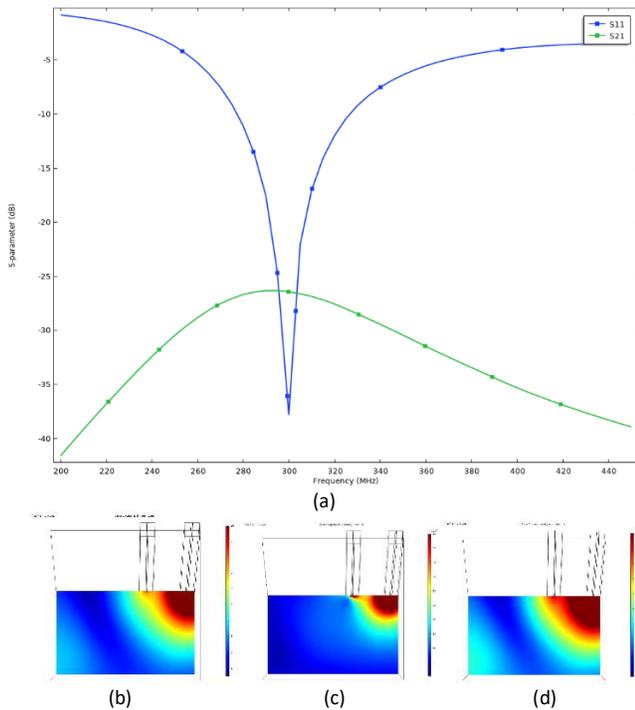

Fig. 4 The decoupling performance of the MTL resonators in a planar configuration. (a) S parameters evaluation (b) Electric fields, (c) magnetic fields, and (d) currents induced in the passive element from the active component.

*C. Magnetic Field Distribution*

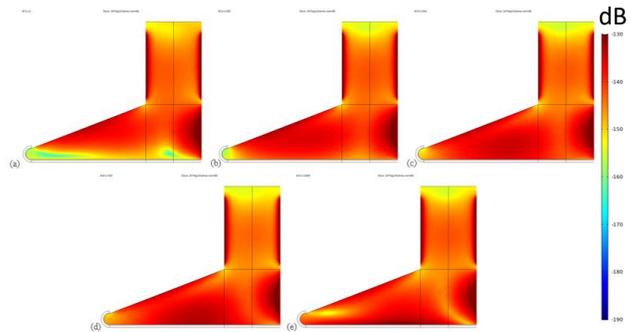

Fig. 5 Magnetic field distribution produced by the proposed 13-channel hybrid array system with (a) No dielectric sheet (b) dielectric sheet, $\varepsilon_r$: 300 (c) dielectric sheet, $\varepsilon_r$: 500 (d) dielectric sheet, $\varepsilon_r$: 700 (e) dielectric sheet, $\varepsilon_r$: 1000

The magnetic field distribution in decibels was shown on a logarithmic scale. The following equation was used: 20*log10 (emw. normB). Our findings indicate that a dielectric layer with a high dielectric constant increases B1 field uniformity. Magnetic field distribution and coverage were significantly improved using dielectric sheets with relative permittivity of $\varepsilon_r$: 500 and 700. However, as the relative permittivity value is increased further to an acceptable value, the findings show that any field cancellation is developing in the region of interest. As a result, determining the dielectric sheet's suitable relative permittivity [101-105] is crucial.

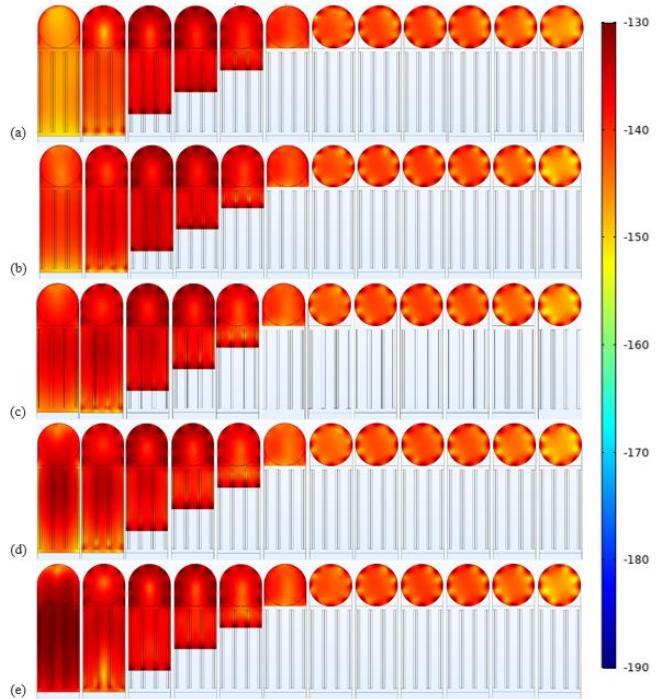

Fig. 6 Axial slices of the Magnetic field distribution in the human foot phantom produced by the proposed 13-channel hybrid array system with (a)

No dielectric sheet (b) dielectric sheet, ε$_r$: 300 (c) dielectric sheet, ε$_r$: 500 (d) dielectric sheet, ε$_r$: 700 (e) dielectric sheet, ε$_r$: 1000 relative permittivity of the dielectric sheet used in the array system.

Acknowledgment

This work is supported in part by the grant from NIH (U01 EB023829) and SUNY Empire Innovation Professorship Award.

## D. Specific Absorption Rate

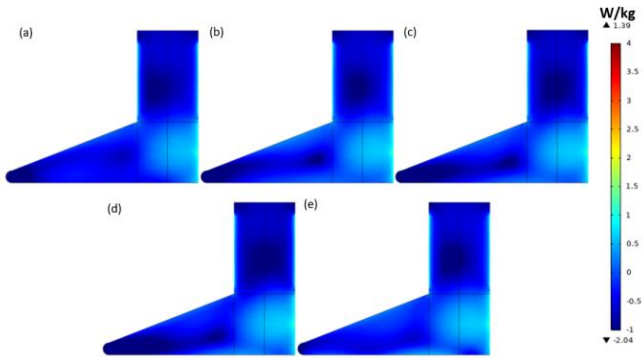

Fig. 7 Specific Absorption Rate map in a sagittal slice of the human foot/ankle-shaped phantom produced by the proposed 13-channel hybrid array system with (a) No dielectric sheet, (b) dielectric sheet, ε$_r$: 300, (c) dielectric sheet, ε$_r$: 500 (d) dielectric sheet, ε$_r$: 700 (e) dielectric sheet, ε$_r$: 1000.

The specific absorption rate induced in the area of interest, the human foot phantom, was evaluated using electromagnetic simulations. The expression used was as follows: (emw. SAR). The local SAR values are under the permissible FDA guideline limits, and hence, the proposed array can be considered safe for Ultrahigh field MR applications.

## IV. CONCLUSION AND DISCUSSION

We designed and simulated a novel RF coil array system for ultra-high field foot/ankle magnetic resonance imaging applications in this study. According to the findings, our designed array system could produce magnetic field distributions that covered the whole region of interest. As capacitively terminated microstrip transmission lines are arranged in an array, they exhibit excellent decoupling efficiency. In addition, the 13-channel hybrid array system's magnetic field coverage and uniformity were increased by using a high dielectric sheet. In the future, Bench tests may be carried out by building array systems and testing their success in a realistic setting. Accurate safety studies could help make more design changes. To improve the field distributions, individual analysis can be performed to precisely calculate the

REFERENCE

[1] P. C. Lauterbur, "Image Formation by Induced Local Interaction: Examples employing Nuclear Magnetic Resonance," *Nature,* vol. 241, pp. 190-191, 1973.

[2] P. Mansfield, "Multi-planar image formation using NMR spin echoes," *J. Phys. C,* vol. 10, pp. L55-L58, 1977.

[3] D. Burstein, A. Bashir, and M. L. Gray, "MRI techniques in early stages of cartilage disease," *Invest Radiol,* vol. 35, no. 10, pp. 622-38, Oct 2000, doi: 10.1097/00004424-200010000-00008.

[4] E. Jimenez-Boj *et al.*, "Bone erosions and bone marrow edema as defined by magnetic resonance imaging reflect true bone marrow inflammation in rheumatoid arthritis," *Arthritis Rheum,* vol. 56, no. 4, pp. 1118-24, Apr 2007, doi: 10.1002/art.22496.

[5] S. Bertram, A. Rudisch, K. Innerhofer, E. Pumpel, G. Grubwieser, and R. Emshoff, "Diagnosing TMJ internal derangement and osteoarthritis with magnetic resonance imaging," *J Am Dent Assoc,* vol. 132, no. 6, pp. 753-61, Jun 2001, doi: 10.14219/jada.archive.2001.0272.

[6] K. N. Malizos, A. H. Zibis, Z. Dailiana, M. Hantes, T. Karachalios, and A. H. Karantanas, "MR imaging findings in transient osteoporosis of the hip," *Eur J Radiol,* vol. 50, no. 3, pp. 238-44, Jun 2004, doi: 10.1016/j.ejrad.2004.01.020.

[7] J. D. Saliman, C. F. Beaulieu, and T. R. McAdams, "Ligament and tendon injury to the elbow: clinical, surgical, and imaging features," *Top Magn Reson Imaging,* vol. 17, no. 5, pp. 327-36, Oct 2006, doi: 10.1097/RMR.0b013e3180421c9e.

[8] M. E. Ladd *et al.*, "Pros and cons of ultra-high-field MRI/MRS for human application," *Prog Nucl Magn Reson Spectrosc,* vol. 109, pp. 1-50, Dec 2018, doi: 10.1016/j.pnmrs.2018.06.001.

[9] R. Krug *et al.*, "Ultrashort echo time MRI of cortical bone at 7 tesla field strength: A feasibility study," *J Magn Reson Imaging,* vol. 34, no. 3, pp. 691-695, Jul 18 2011. [Online]. Available: http://www.ncbi.nlm.nih.gov/entrez/query.fcgi?cmd=Retrieve&db=PubMed&dopt=Citation&list_uids=21769960

[10] Y. Pang and X. Zhang, "Interpolated Compressed Sensing MR Image Reconstruction using Neighboring Slice k-space Data," in *Proceedings of the 20th Annual Meeting of ISMRM, Melbourne, Australia*, 2012, p. 2275.

[11] Y. Pang and X. Zhang, "Interpolated compressed sensing for 2D multiple slice fast MR imaging," *PLoS One,* vol. 8, no. 2, p. e56098, 2013. [Online]. Available: http://www.ncbi.nlm.nih.gov/entrez/query.fcgi?cmd=Retrieve&db=PubMed&dopt=Citation&list_uids=23409130

[12] X. Zhang *et al.*, "Human extremity imaging using microstrip resonators at 7T," in *Proceedings of the 21st Annual Meeting of ISMRM, Salt Lake City, USA*, 2013, p. 1675.

[13] B. Wu *et al.*, "7T human Spine Arrays with Adjustable Inductive Decoupling," in *Proceedings of the 17th Annual Meeting of ISMRM, Honolulu, USA*, 2009, p. 2997.

[14] B. Wu *et al.*, "7T human spine imaging arrays with adjustable inductive decoupling," *IEEE Trans Biomed Eng,* vol. 57, no. 2, pp. 397-403, Feb 2010. [Online]. Available: http://www.ncbi.nlm.nih.gov/entrez/query.fcgi?cmd=Retrieve&db=PubMed&dopt=Citation&list_uids=19709956

[15] C. Wang, A. Pai, B. Wu, R. Krug, S. Majumdar, and X. Zhang, "A Microstrip Volume Coil with Easy Access for Wrist Imaging at 7T," in *Proceedings of the 17th Annual Meeting of ISMRM, Honolulu, USA*, 2009, p. 2946.

[16] G. Aringhieri, V. Zampa, and M. Tosetti, "Musculoskeletal MRI at 7 T: do we need more or is it more than enough?," *Eur Radiol Exp,* vol. 4, no. 1, p. 48, Aug 6 2020, doi: 10.1186/s41747-020-00174-1.

[17] G. Dean Deyle, "The role of MRI in musculoskeletal practice: a clinical perspective," *J Man Manip Ther,* vol. 19, no. 3, pp. 152-61, Aug 2011, doi: 10.1179/2042618611Y.0000000009.

[18] J. E. Sage and P. Gavin, "Musculoskeletal MRI," *Vet Clin North Am Small Anim Pract,* vol. 46, no. 3, pp. 421-51, v, May 2016, doi: 10.1016/j.cvsm.2015.12.003.

[19] M. Tafur, Z. S. Rosenberg, and J. T. Bencardino, "MR Imaging of the Midfoot Including Chopart and Lisfranc Joint Complexes," *Magn Reson Imaging Clin N Am,* vol. 25, no. 1, pp. 95-125, Feb 2017, doi: 10.1016/j.mric.2016.08.006.

[20] L. P. Liu *et al.*, "Diagnostic Performance of Diffusion-weighted Magnetic Resonance Imaging in Bone Malignancy: Evidence From a Meta-Analysis," *Medicine (Baltimore),* vol. 94, no. 45, p. e1998, Nov 2015, doi: 10.1097/MD.0000000000001998.

[21] S. Qi *et al.*, "Cortical inhibition deficits in recent onset PTSD after a single prolonged trauma

exposure," *Neuroimage Clin,* vol. 3, pp. 226-33, 2013, doi: 10.1016/j.nicl.2013.08.013.

[22] J. Lee et al., "Open-chest 31P magnetic resonance spectroscopy of mouse heart at 4.7 Tesla," *J Magn Reson Imaging,* vol. 24, no. 6, pp. 1269-76, Dec 2006, doi: 10.1002/jmri.20766.

[23] X. Hu, Y. Li, L. Zhang, X. Zhang, X. Liu, and Y. C. Chung, "A 32-channel coil system for MR vessel wall imaging of intracranial and extracranial arteries at 3T," *Magn Reson Imaging,* vol. 36, pp. 86-92, Feb 2017, doi: 10.1016/j.mri.2016.10.018.

[24] S. Qi et al., "Anomalous gray matter structural networks in recent onset post-traumatic stress disorder," *Brain Imaging Behav,* vol. 12, no. 2, pp. 390-401, Apr 2018, doi: 10.1007/s11682-017-9693-z.

[25] Y. X. Wang, G. G. Lo, J. Yuan, P. E. Larson, and X. Zhang, "Magnetic resonance imaging for lung cancer screen," *J Thorac Dis,* vol. 6, no. 9, pp. 1340-8, Sep 2014, doi: 10.3978/j.issn.2072-1439.2014.08.43.

[26] Y. Li and X. Zhang, "Advanced MR Imaging Technologies in Fetuses," *OMICS J Radiol,* vol. 1, no. 4, p. e113, Sep 2012, doi: 10.4172/2167-7964.1000e113.

[27] X. Zhang et al., "High Resolution Imaging of the Human Head at 8 Tesla," in *Proceedings of ESMRMB annual meeting, Sevilla, Spain*, Sevilla, Spain, 1999, p. 44.

[28] D. I. Hoult, "Sensitivity and power deposition in a high-field imaging experiment," *J Magn Reson Imaging,* vol. 12, no. 1, pp. 46-67., 2000.

[29] K. Ugurbil et al., "Magnetic resonance studies of brain function and neurochemistry," *Annu Rev Biomed Eng,* vol. 2, pp. 633-660, 2000.

[30] G. Adriany et al., "Transmit and receive transmission line arrays for 7 Tesla parallel imaging," *Magn Reson Med,* vol. 53, no. 2, pp. 434-45, Feb 2005. [Online]. Available: http://www.ncbi.nlm.nih.gov/entrez/query.fcgi?cmd=Retrieve&db=PubMed&dopt=Citation&list_uids=15678527

[31] H. Lei, X. H. Zhu, X. L. Zhang, K. Ugurbil, and W. Chen, "In vivo 31P magnetic resonance spectroscopy of human brain at 7 T: an initial experience," *Magn Reson Med,* vol. 49, no. 2, pp. 199-205, Feb 2003. [Online]. Available: http://www.ncbi.nlm.nih.gov/entrez/query.fcgi?cmd=Retrieve&db=PubMed&dopt=Citation&list_uids=12541238.

[32] X. H. Zhu et al., "Quantitative imaging of energy expenditure in human brain," *Neuroimage,* vol. 60, no. 4, pp. 2107-17, May 1 2012, doi: 10.1016/j.neuroimage.2012.02.013.

[33] J. Kurhanewicz et al., "Hyperpolarized (13)C MRI: Path to Clinical Translation in Oncology," *Neoplasia,* vol. 21, no. 1, pp. 1-16, Jan 2019, doi: 10.1016/j.neo.2018.09.006.

[34] X. H. Zhu et al., "Development of (17)O NMR approach for fast imaging of cerebral metabolic rate of oxygen in rat brain at high field," *Proc Natl Acad Sci U S A,* vol. 99, no. 20, pp. 13194-9, Oct 1 2002, doi: 10.1073/pnas.202471399.

[35] J. Zhang et al., "Decreased gray matter volume in the left hippocampus and bilateral calcarine cortex in coal mine flood disaster survivors with recent onset PTSD," *Psychiatry Res,* vol. 192, no. 2, pp. 84-90, May 31 2011, doi: 10.1016/j.pscychresns.2010.09.001.

[36] F. Du et al., "In vivo evidence for cerebral bioenergetic abnormalities in schizophrenia measured using 31P magnetization transfer spectroscopy," *JAMA Psychiatry,* vol. 71, no. 1, pp. 19-27, Jan 2014, doi: 10.1001/jamapsychiatry.2013.2287.

[37] H. Qiao, X. Zhang, X. H. Zhu, F. Du, and W. Chen, "In vivo 31P MRS of human brain at high/ultrahigh fields: a quantitative comparison of NMR

detection sensitivity and spectral resolution between 4 T and 7 T," *Magn Reson Imaging,* vol. 24, no. 10, pp. 1281-6, Dec 2006, doi: 10.1016/j.mri.2006.08.002.

[38] X. H. Zhu *et al.*, "Advanced In Vivo Heteronuclear MRS Approaches for Studying Brain Bioenergetics Driven by Mitochondria," *Methods Mol Biol,* vol. 489, pp. 317-57, 2009, doi: 10.1007/978-1-59745-543-5_15.

[39] X. Zhang, X. H. Zhu, R. Tian, Y. Zhang, H. Merkle, and W. Chen, "Measurement of arterial input function of 17O water tracer in rat carotid artery by using a region-defined (REDE) implanted vascular RF coil," *MAGMA,* vol. 16, no. 2, pp. 77-85, Jul 2003, doi: 10.1007/s10334-003-0013-9.

[40] H. Zhu *et al.*, "High-field 17O study of 3D CMRO2 imaging in human visual cortex," in *Proceedings of the 14th Annual Meeting of ISMRM, Seattle, USA*, 2006, p. 409.

[41] X. Zhu, X. Zhang, S. Tang, S. Ogawa, K. Ugurbil, and W. Chen, "Probing fast neuronal interaction in the human ocular dominate columns based on fMRI BOLD response at 7 Tesla," in *Proceedings of the 9th Annual Meeting of ISMRM, Glasgow, Scotland*, 2001.

[42] X. Zhu, X. Zhang, and W. Chen, "Study of 17O NMR sensitivity and relaxation times of cerebral water in human at 7 Tesla," in *Proceedings of the 11th Annual Meeting of ISMRM, Toronto, Canada*, Toronto, Canada, 2003, p. 868.

[43] E. Milshteyn and X. Zhang, "The Need and Initial Practice of Parallel Imaging and Compressed Sensing in Hyperpolarized (13)C MRI in vivo," *OMICS J Radiol,* vol. 4, no. 4, 2015, doi: 10.4172/2167-7964.1000e133.

[44] X. Zhang, K. Ugurbil, and W. Chen, "Microstrip RF surface coil design for extremely high-field MRI and spectroscopy," *Magn Reson Med,* vol. 46, no. 3, pp. 443-50, Sep 2001, doi: 10.1002/mrm.1212.

[45] X. Zhang, K. Ugurbil, R. Sainati, and W. Chen, "An inverted-microstrip resonator for human head proton MR imaging at 7 tesla," *IEEE Trans Biomed Eng,* vol. 52, no. 3, pp. 495-504, Mar 2005, doi: 10.1109/TBME.2004.842968.

[46] X. Zhang, X. H. Zhu, and W. Chen, "Higher-order harmonic transmission-line RF coil design for MR applications," (in eng), *Magn Reson Med,* vol. 53, no. 5, pp. 1234-9, May 2005, doi: 10.1002/mrm.20462.

[47] Q. X. Yang *et al.*, "Analysis of wave behavior in lossy dielectric samples at high field," *Magn Reson Med,* vol. 47, no. 5, pp. 982-9, May 2002. [Online]. Available: http://www.ncbi.nlm.nih.gov/entrez/query.fcgi?cmd=Retrieve&db=PubMed&dopt=Citation&list_uids=11979578.

[48] Q. X. Yang *et al.*, "Phantom design method for high-field MRI human systems," *Magn Reson Med,* vol. 52, no. 5, pp. 1016-20, Nov 2004. [Online]. Available: http://www.ncbi.nlm.nih.gov/entrez/query.fcgi?cmd=Retrieve&db=PubMed&dopt=Citation&list_uids=15508165

[49] Q. X. Yang *et al.*, "Manipulation of image intensity distribution at 7.0 T: passive RF shimming and focusing with dielectric materials," *J Magn Reson Imaging,* vol. 24, no. 1, pp. 197-202, Jul 2006. [Online]. Available: http://www.ncbi.nlm.nih.gov/entrez/query.fcgi?cmd=Retrieve&db=PubMed&dopt=Citation&list_uids=16755543

[50] B. Wu, Y. Li, C. Wang, D. B. Vigneron, and X. Zhang, "Multi-reception strategy with improved SNR for multichannel MR imaging," *PLoS One,* vol. 7, no. 8, p. e42237, 2012. [Online]. Available: http://www.ncbi.nlm.nih.gov/entrez/query.fcgi?cmd=Retrieve&db=PubMed&dopt=Citation&list_uids=22879921

[51] B. Wu *et al.*, "Shielded microstrip array for 7T human MR imaging," *IEEE Trans Med Imaging,* vol. 29, no. 1, pp. 179-84, Jan 2010. [Online]. Available: http://www.ncbi.nlm.nih.gov/entrez/query.fcgi?cmd=Retrieve&db=PubMed&dopt=Citation&list_uids=19822470

[52] B. Wu *et al.*, "Acquisition of 7T Human Spine Imaging," in *Proceedings of the 17th Annual Meeting of ISMRM, Honolulu, USA*, 2009, p. 3187.

[53] Y. Pang and X. Zhang, "Multi-voxel Excitation using Compressed Sensing Typed Sparse Pulse," in *the 18th Annual meeting of ISMRM* Stockholm, Sweden, 2010, p. 4952.

[54] Y. Pang and X. Zhang, "Sparse Parallel Transmission using Optimized Sparse k-space trajectory by Simulated Annealing," in *the 18th Annual Meeting of ISMRM*, Stockholm, Sweden, 2010, p. 4921.

[55] Y. Pang, B. Wu, C. Wang, D. B. Vigneron, and X. Zhang, "Numerical Analysis of Human Sample Effect on RF Penetration and Liver MR Imaging at Ultrahigh Field," *Concepts Magn Reson Part B Magn Reson Eng,* vol. 39B, no. 4, pp. 206-216, Oct 2011. [Online]. Available: http://www.ncbi.nlm.nih.gov/entrez/query.fcgi?cmd=Retrieve&db=PubMed&dopt=Citation&list_uids=22337345

[56] Y. Pang, G. X. Shen, and X. Zhang, "Two dimensional spatial selective Shinnar Le Roux pulse design for arbitrary k-trajectory," in *the 17th Annual meeting of ISMRM*, Honolulu, HI, 2009, p. 2598.

[57] E. Moser, "Ultra-high-field magnetic resonance: Why and when?," *World J Radiol,* vol. 2, no. 1, pp. 37-40, Jan 28 2010, doi: 10.4329/wjr.v2.i1.37.

[58] Y. Li, C. Wang, B. Yu, D. Vigneron, W. Chen, and X. Zhang, "Image homogenization using pre-emphasis method for high field MRI," *Quant Imaging Med Surg,* vol. 3, no. 4, pp. 217-23, Aug 2013. [Online]. Available: http://www.ncbi.nlm.nih.gov/entrez/query.fcgi?cmd=Retrieve&db=PubMed&dopt=Citation&list_uids=24040618

[59] Q. Yang, H. Zhang, J. Xia, and X. Zhang, "Evaluation of magnetic resonance image segmentation in brain low-grade gliomas using support vector machine and convolutional neural network," *Quant Imaging Med Surg,* vol. 11, no. 1, pp. 300-316, Jan 2021, doi: 10.21037/qims-20-783.

[60] X. Zhang and J. X. Ji, "Parallel and sparse MR imaging: methods and instruments-Part 1," *Quant Imaging Med Surg,* vol. 4, no. 1, pp. 1-3, Feb 2014, doi: 10.3978/j.issn.2223-4292.2014.03.01.

[61] J. X. Ji and X. Zhang, "Parallel and sparse MR imaging: methods and instruments-Part 2," *Quant Imaging Med Surg,* vol. 4, no. 2, pp. 68-70, Apr 2014, doi: 10.3978/j.issn.2223-4292.2014.04.16.

[62] Y. Pang, B. Yu, and X. Zhang, "Enhancement of the low resolution image quality using randomly sampled data for multi-slice MR imaging," *Quant Imaging Med Surg,* vol. 4, no. 2, pp. 136-44, Apr 2014, doi: 10.3978/j.issn.2223-4292.2014.04.17.

[63] S. O. Dumoulin, A. Fracasso, W. van der Zwaag, J. C. W. Siero, and N. Petridou, "Ultra-high field MRI: Advancing systems neuroscience towards mesoscopic human brain function," *Neuroimage,* vol. 168, pp. 345-357, Mar 2018, doi: 10.1016/j.neuroimage.2017.01.028.

[64] C. Wang and G. X. Shen, "B1 field, SAR, and SNR comparisons for birdcage, TEM, and microstrip coils at 7T," *J Magn Reson Imaging,* vol. 24, no. 2, pp. 439-43, Aug 2006, doi: 10.1002/jmri.20635.

[65] X. Yan, Z. Cao, and X. Zhang, "Simulation verification of SNR and parallel imaging improvements by ICE-decoupled loop array in MRI," *Appl Magn Reson,* vol. 47, no. 4, pp. 395-403, Apr 2016, doi: 10.1007/s00723-016-0764-x.

[66] Y. Pang, B. Wu, X. Jiang, D. B. Vigneron, and X. Zhang, "Tilted microstrip phased arrays with improved electromagnetic decoupling for ultrahigh-field magnetic resonance imaging," *Medicine (Baltimore),* vol. 93, no. 28, p. e311, Dec 2014, doi: 10.1097/MD.0000000000000311.

[67] C. M. Collins *et al.*, "Different excitation and reception distributions with a single-loop transmit-receive surface coil near a head-sized spherical phantom at 300 MHz," *Magn Reson Med,* vol. 47, no. 5, pp. 1026-8, May 2002. [Online]. Available: http://www.ncbi.nlm.nih.gov/entrez/query.fcgi?cmd=Retrieve&db=PubMed&dopt=Citation&list_uids=11979585

[68] J. Wang *et al.*, "Polarization of the RF field in a human head at high field: a study with a quadrature surface coil at 7.0 T," *Magn Reson Med,* vol. 48, no. 2, pp. 362-9, Aug 2002. [Online]. Available: http://www.ncbi.nlm.nih.gov/entrez/query.fcgi?cmd=Retrieve&db=PubMed&dopt=Citation&list_uids=12210945.

[69] X. Zhang, "Experimental Design of Transmission Line Volume RF coil for MR Imaging at 8T," in *Proceedings of the 8th Annual Meeting of ISMRM, Denver, USA*, 2000, p. 150.

[70] X. Zhang *et al.*, "Design of catheter radio frequency coils using coaxial transmission line resonators for interventional neurovascular MR imaging," *Quant Imaging Med Surg,* vol. 7, no. 2, pp. 187-194, Apr 2017, doi: 10.21037/qims.2016.12.05.

[71] Y. Li, Y. Pang, and X. Zhang, "Common-mode differential-mode (CMDM) method for quadrature transmit/receive surface coil for ultrahigh field MRI," in *the 19th Annual Meeting of ISMRM*, Montreal, Canada, 2011.

[72] Y. Li, Z. Xie, Y. Pang, D. Vigneron, and X. Zhang, "ICE decoupling technique for RF coil array designs," *Med Phys,* vol. 38, no. 7, pp. 4086-93, Jul 2011, doi: 10.1118/1.3598112.

[73] J. Lu, Y. Pang, C. Wang, B. Wu, D. B. Vigneron, and X. Zhang, "Evaluation of Common RF Coil Setups for MR Imaging at Ultrahigh Magnetic Field: A Numerical Study," *Int Symp Appl Sci Biomed Commun Technol,* vol. 2011, 2011, doi: 10.1145/2093698.2093768.

[74] Y. Pang, D. B. Vigneron, and X. Zhang, "Parallel traveling-wave MRI: a feasibility study," *Magn Reson Med,* vol. 67, no. 4, pp. 965-78, Apr 2012. [Online]. Available: http://www.ncbi.nlm.nih.gov/entrez/query.fcgi?cmd=Retrieve&db=PubMed&dopt=Citation&list_uids=21858863

[75] Y. Pang, E. W. Wong, B. Yu, and X. Zhang, "Design and numerical evaluation of a volume coil array for parallel MR imaging at ultrahigh fields," *Quant Imaging Med Surg,* vol. 4, no. 1, pp. 50-6, Feb 2014, doi: 10.3978/j.issn.2223-4292.2014.02.07.

[76] Y. Pang, Z. Xie, Y. Li, D. Xu, D. Vigneron, and X. Zhang, "Resonant Mode Reduction in Radiofrequency Volume Coils for Ultrahigh Field Magnetic Resonance Imaging," *Materials (Basel),* vol. 4, no. 8, pp. 1333-1344, Jul 28 2011. [Online]. Available: http://www.ncbi.nlm.nih.gov/entrez/query.fcgi?cmd=Retrieve&db=PubMed&dopt=Citation&list_uids=22081791

[77] Y. Pang *et al.*, "A dual-tuned quadrature volume coil with mixed lambda/2 and lambda/4 microstrip resonators for multinuclear MRSI at 7 T," *Magn Reson Imaging,* vol. 30, no. 2, pp. 290-8, Feb 2012. [Online]. Available: http://www.ncbi.nlm.nih.gov/entrez/query.fcgi?cmd=Retrieve&db=PubMed&dopt=Citation&list_uids=22055851

[78] Y. Pang, B. Yu, D. B. Vigneron, and X. Zhang, "Quadrature transmit array design using single-

feed circularly polarized patch antenna for parallel transmission in MR imaging," *Quant Imaging Med Surg,* vol. 4, no. 1, pp. 11-8, Feb 2014, doi: 10.3978/j.issn.2223-4292.2014.02.03.

[79] Y. Pang and X. Zhang, "Precompensation for mutual coupling between array elements in parallel excitation," *Quant Imaging Med Surg,* vol. 1, no. 1, pp. 4-10, Dec 2011. [Online]. Available: http://www.ncbi.nlm.nih.gov/entrez/query.fcgi?cmd=Retrieve&db=PubMed&dopt=Citation&list_uids=23243630

[80] C. Wang *et al.*, "A practical multinuclear transceiver volume coil for in vivo MRI/MRS at 7 T," *Magn Reson Imaging,* vol. 30, no. 1, pp. 78-84, Jan 2012. [Online]. Available: http://www.ncbi.nlm.nih.gov/entrez/query.fcgi?cmd=Retrieve&db=PubMed&dopt=Citation&list_uids=22055858

[81] X. Zhang, K. Ugurbil, and W. Chen, "Method and apparatus for magnetic resonance imaging and spectroscopy using microstrip transmission line coils," (in English), Patent 7023209 Patent Appl. 09/974,184, 2006.

[82] X. Zhang, "Method and apparatus for MRI signal excitation and reception using non-resonance RF method (NORM)," (in English), USA Patent 8269498 Patent Appl. 12/533,724, Sep. 18, 2012, 2012.

[83] C. Wang and X. Zhang, "Evaluation of B1+ and E field of RF Resonator with High Dielectric Insert," in *Proceedings of the 17th Annual Meeting of ISMRM, Honolulu, Hawaii*, 2009, p. 3054.

[84] M. S. H. Akram, T. Obata, and T. Yamaya, "Microstrip Transmission Line RF Coil for a PET/MRI Insert," *Magn Reson Med Sci,* vol. 19, no. 2, pp. 147-153, May 1 2020, doi: 10.2463/mrms.mp.2019-0137.

[85] X. Zhang, K. Ugurbil, and W. Chen, "A microstrip transmission line volume coil for human head MR imaging at 4T," *J Magn Reson,* vol. 161, no. 2, pp. 242-51, Apr 2003, doi: 10.1016/s1090-7807(03)00004-1.

[86] X. Zhang, "Novel radio frequency resonators for in vivo magnetic resonance imaging and spectroscopy at very high magnetic fields," Ph.D., Biomedical Engineering, University of Minnesota, 2002.

[87] X. Zhang, K. Ugurbil, and W. Chen, "Method and apparatus for magnetic resonance imaging and spectroscopy using microstrip transmission line coils," Patent Appl. 11224436, 2005.

[88] X. Zhang, K. Ugurbil, and W. Chen, "Method and apparatus for magnetic resonance imaging and spectroscopy using microstrip transmission line coils. 7023209," *US patent,* 2006.

[89] X. Zhang, K. Ugurbil, and W. Chen, "Microstrip RF Surface Coils for Human MRI Studies at 7 Tesla," in *Proceedings of the 9th Annual Meeting of ISMRM, Glasgow, Scotland*, 2001, p. 1104.

[90] X. Zhang, K. Ugurbil, and W. Chen, "A New RF Volume Coil for Human MR Imaging and Spectroscopy at 4 Tesla," in *Proceedings of the 9th Annual Meeting of ISMRM, Glasgow, Scotland*, Glasgow, Scotland, 2001, p. 1103.

[91] X. Zhang, R. Sainati, T. Vaughan, and W. Chen, "Analysis of single microstrip resonator with capacitive termination at very high fields," in *9th ISMRM meeting*, UK, 2001, vol. Submitted.

[92] X. Zhang and Z. Xie, "Method and apparatus for magnetic resonance imaging and spectroscopy using multiple-mode coils," Patent Appl. 12/990,541, 2009.

[93] C. Tian, A. W. Magill, A. Comment, R. Gruetter, and L. Hongxia, "Ultra-high field birdcage coils: a comparison study at 14.1T," *Annu Int Conf IEEE Eng Med Biol Soc,* vol. 2014, pp. 2360-3, 2014, doi: 10.1109/EMBC.2014.6944095.

[94] S. F. Ahmad, Y. C. Kim, I. C. Choi, and H. D. Kim, "Recent Progress in Birdcage RF Coil Technology for MRI System," *Diagnostics (Basel),* vol. 10, no. 12, Nov 27 2020, doi: 10.3390/diagnostics10121017.

[95] P. Heo *et al.*, "Multi-port-driven birdcage coil for multiple-mouse MR imaging at 7 T," *Scanning,* vol. 38, no. 6, pp. 747-756, Nov 2016, doi: 10.1002/sca.21324.

[96] N. Lopez Rios *et al.*, "Design and construction of an optimized transmit/receive hybrid birdcage resonator to improve full body images of medium-sized animals in 7T scanner," *PLoS One,* vol. 13, no. 2, p. e0192035, 2018, doi: 10.1371/journal.pone.0192035.

[97] L. I. Navarro de Lara, L. G. Rad, S. N. Makarov, J. Stockmann, L. L. Wald, and A. Nummenmaa, "Simulations of a birdcage coil B1+ field on a human body model for designing a 3T multichannel TMS/MRI head coil array," *Annu Int Conf IEEE Eng Med Biol Soc,* vol. 2018, pp. 4752-4755, Jul 2018, doi: 10.1109/EMBC.2018.8513208.

[98] M. Tang, K. Okamoto, T. Haruyama, and T. Yamamoto, "Electromagnetic simulation of RF burn injuries occurring at skin-skin and skin-bore wall contact points in an MRI scanner with a birdcage coil," *Phys Med,* vol. 82, pp. 219-227, Feb 2021, doi: 10.1016/j.ejmp.2021.02.008.

[99] C. L. Chin, C. M. Collins, S. Li, B. J. Dardzinski, and M. B. Smith, "BirdcageBuilder: Design of Specified-Geometry Birdcage Coils with Desired Current Pattern and Resonant Frequency," *Concepts Magn Reson,* vol. 15, no. 2, pp. 156-163, Jun 2002, doi: 10.1002/cmr.10030.

[100] D. E. Vincent, T. Wang, T. A. K. Magyar, P. I. Jacob, R. Buist, and M. Martin, "Birdcage volume coils and magnetic resonance imaging: a simple experiment for students," *J Biol Eng,* vol. 11, p. 41, 2017, doi: 10.1186/s13036-017-0084-1.

[101] W. M. Teeuwisse, W. M. Brink, K. N. Haines, and A. G. Webb, "Simulations of high permittivity materials for 7 T neuroimaging and evaluation of a new barium titanate-based dielectric," *Magn Reson Med,* vol. 67, no. 4, pp. 912-8, Apr 2012, doi: 10.1002/mrm.24176.

[102] T. P. A. O'Reilly, A. G. Webb, and W. M. Brink, "Practical improvements in the design of high permittivity pads for dielectric shimming in neuroimaging at 7T," *J Magn Reson,* vol. 270, pp. 108-114, Sep 2016, doi: 10.1016/j.jmr.2016.07.003.

[103] K. Haines, N. B. Smith, and A. G. Webb, "New high dielectric constant materials for tailoring the B1+ distribution at high magnetic fields," *J Magn Reson,* vol. 203, no. 2, pp. 323-7, Apr 2010, doi: 10.1016/j.jmr.2010.01.003.

[104] J. E. Snaar *et al.*, "Improvements in high-field localized MRS of the medial temporal lobe in humans using new deformable high-dielectric materials," *NMR Biomed,* vol. 24, no. 7, pp. 873-9, Aug 2011, doi: 10.1002/nbm.1638.

[105] W. M. Brink and A. G. Webb, "High permittivity pads reduce specific absorption rate, improve B1 homogeneity, and increase contrast-to-noise ratio for functional cardiac MRI at 3 T," *Magn Reson Med,* vol. 71, no. 4, pp. 1632-40, Apr 2014, doi: 10.1002/mrm.24778.